\documentclass[sigconf]{acmart}
\usepackage{subcaption} 
\AtBeginDocument{%
  \providecommand\BibTeX{{%
    \normalfont B\kern-0.5em{\scshape i\kern-0.25em b}\kern-0.8em\TeX}}}

\setcopyright{rightsretained}
\copyrightyear{2024}
\acmYear{2024}
\acmDOI{}

\acmConference{}
\acmPrice{}
\acmISBN{}

\begin{document}

\title[Window Operation Detection]{Towards a data-driven and scalable approach for window operation detection in multi-family residential buildings}

\author{Juliet Nwagwu Ume-Ezeoke}
\affiliation{%
  \institution{Stanford University}
  \city{Stanford}
  \state{CA}
  \country{USA}
  \postcode{94305}
}

\author{Kopal Nihar}
\affiliation{%
  \institution{Stanford University}
  \city{Stanford}
  \state{CA}
  \country{USA}
  \postcode{94305}
}

\author{Catherine Gorle}
\affiliation{%
  \institution{Stanford University}
  \city{Stanford}
  \state{CA}
  \country{USA}
  \postcode{94305}
}

\author{Rishee Jain}
\affiliation{%
  \institution{Stanford University}
  \city{Stanford}
  \state{CA}
  \country{USA}
  \postcode{94305}
}

\renewcommand{\shortauthors}{Nwagwu Ume-Ezeoke, et al.}

\begin{abstract}
Natural cooling, utilizing non-mechanical cooling, presents a low-carbon and low-cost way to provide thermal comfort in residential buildings. However, designing naturally cooled buildings requires a clear understanding of how opening and closing windows affect occupants' comfort. Predicting when and why occupants open windows is a challenging task, often relying on specialized sensors and building-specific training data. This limits the scalability of natural cooling solutions. Here, we propose a novel unsupervised method that utilizes easily deployable off-the-shelf temperature and humidity sensors to detect window operations. The effectiveness of our approach is evaluated using an empirical dataset and compared with a state-of-the-art support vector machine (SVM) model. The results demonstrate that our proposed method outperforms the SVM on key indicators, except when indoor and outdoor temperatures have small differences. Unlike the SVM’s sensitivity to time series characteristics, our proposed method relies solely on indoor temperature and exhibits robust performance in pilot studies, making it a promising candidate for developing a highly scalable and generalizable window operation detection model. This work demonstrates the potential of unsupervised data-driven methods for understanding window operations in residential buildings. By enabling more accurate modeling of naturally cooled buildings, our work aims to facilitate the widespread adoption of this low-cost and low-carbon technology.
\end{abstract}



\maketitle

\section{Introduction}

Over the last few decades, building-related greenhouse gas emissions have surged despite efforts to improve energy efficiency. This surge is accelerated by widespread adoption of air-conditioning systems which are projected to increase from approximately 1.2 million units to a staggering 4.5 billion units by 2050, according to an IEA report in 2018 \cite{iea_future_2018}. In this context, alternative cooling strategies, such as natural cooling, offer a low-energy and low-cost pathway to achieve thermal comfort. Natural cooling possesses the potential to be resilient, as it can reduce reliance on vulnerable mechanical cooling systems that may fail during heat waves \cite{ahmed_natural_2021}.

However, the design of effective naturally cooled buildings faces challenges arising from the unpredictability of changing weather conditions and the diverse window operating behaviors of occupants \cite{rijal_development_2008}. While significant progress has been made in advancing the robustness of meteorological models, models of building-occupant interaction are still in their infancy \cite{hawila_occupants_2023}. \cite{doca_data-mining_2014} has most notably used predictions of window state based on various parameters to infer patterns of occupant behavior. This work relies on supervised training of a window detection model, whose truth values are often gathered through specialized sensors or self-reporting. However such supervised models have limited generalizability beyond the specific context of data collection, (building type or climate), raising the need for unsupervised models that can infer patterns from pre-existing data. Additionally, tight building design timelines do not easily permit data collection for extended periods of time (6 months or more for previous window detection models), highlighting a critical gap in the development of window detection methods to date.

Therefore, we propose a novel unsupervised window opening and closing detection method that utilizes easy-to-deploy off-the-shelf temperature and humidity sensors and assess our model on a series of short-term datasets. We also compare the effectiveness of our proposed method with a support vector machine model (SVM). Logistic regression \cite{kim_automatic_2019} and Markov chain models \cite{haldi_interactions_2009} have been historically popular for window detection due to their straightforward implementation, while neural network models \cite{markovic_window_2018} more recently been lauded for their high accuracy. Support vector machines (SVMs) \cite{de_rautlin_de_la_roy_deep_2023}  form an ideal middle ground as they are resource-efficient, easy to implement, and demonstrate robust performance across various features. 

In order to validate our proposed method, we not only leverage existing metrics for binary classification such as F-1 score \cite{haldi_interactions_2009}  but domain-specific metrics such as total time of true openings \cite{de_rautlin_de_la_roy_deep_2023}. We also introduce custom metrics to better understand the dynamics of window state changes in an unsupervised setting, in lieu of changing environmental parameters. Overall, our research aims to overcome existing model challenges by leveraging unsupervised data-driven techniques and offers a scalable solution for understanding and optimizing window operations in multi-family residential buildings.

\hypertarget{methods}{%
\section{Methods}\label{methods}}


\hypertarget{data-collection}{%
\subsection{Data Collection}\label{data-collection}}

We collected data for short periods of time over three months in the summer. We placed HOBO® Temp-RH 2.5\% Data Logger (UX100-011) 
sensors in two adjacent rooms in a multi-family residential building,
and measured indoor temperatures, \(T_{meas}(t)\) and relative humidity,
\(RH_{meas}(t)\). In hopes of capturing the average room temperature,
sensors were intentionally positioned away from drafts that could come in
through windows \cite{chen_full-scale_2022}. We also collected data on the ambient
temperature, \(T_{amb}(t)\) and relative humidity \(RH_{amb}(t)\) from a
local weather station . These variables have been shown by \cite{hawila_occupants_2023} to be effective predictors of window state. In one room, the window state, \(W(t)\),
was held constant, while the adjacent room, window state was allowed to vary between open and closed. 

Descriptions of the data in the rooms where the window state varied are
provided in Table~\ref{tbl-data-collected}. The experiments are labeled
as experiments A, B, and C, corresponding to data recorded at the
beginning on July 20, July 27, and September 8 respectively. Experiment
B was much longer than the other experiments, spanning 14 days. Data for
experiments A and C were collected for less than 4 days. 

The three recorded experiments present unique challenges for a detection algorithm. Experiment A presents a
favorable case for window state detection, as it has a fairly even
amount of opening and closed states. Experiment B is unfavorable in this respect since windows are open for 95\% of the recorded time. However, results from a statistical t-test showed that experiment B had the greatest distinction between indoor and ambient temperature t(1388)=47.91, p<0.05, which could be helpful for detection. Experiment C had the smallest distinction t(289)=18.82, p<0.05, as well as an unfavorable balance of window states. 

\hypertarget{tbl-data-collected}{}
\begin{table}
\caption{Data Collected}
\label{tbl-data-collected}
\begin{tabular}{llll}
\toprule
{} &       A &        B &       C \\
\midrule
Starting Day                   &  Jul 20 &   Jul 27 &  Sep 08 \\
Days Recorded                    &  4 &  14 &  3 \\
Opening Percentage             &    58.7 &     94.8 &    77.9 \\
Hours Open                     &    56.5 &    329.0 &    56.2 \\
\bottomrule
\end{tabular}
\end{table}

\hypertarget{detection-methods}{%
\subsection{Detection Methods}\label{detection-methods}}


\hypertarget{new-method-smoothing-technique}{%
\subsubsection{Proposed Method: Smoothing
Technique (ST)}\label{new-method-smoothing-technique}}

Our approach is informed by an understanding of window operations and the underlying thermodynamics principles. 
We expect a typical time series recording of a quantity of
interest, in this case the measured indoor temperature, \(T_{meas}(t)\),
to contain information that reflects the seasonality of ambient
quantities, noise due to occurrences within the space where measurements
are being taken, and the desired signal of changes in the window state. The
way in which these three components of measurement are combined is
unknown, and, the noise component in particular cannot be known based on
the measurements we have collected. As we have recorded information
about the ambient temperature, \(T_{amb}(t)\), we focus on removing the
seasonality from \(T_{meas}(t)\), hypothesizing that this will reveal
the window state signal and noise.


\begin{figure}[h]
\centering
\includegraphics[width=0.3\textwidth]{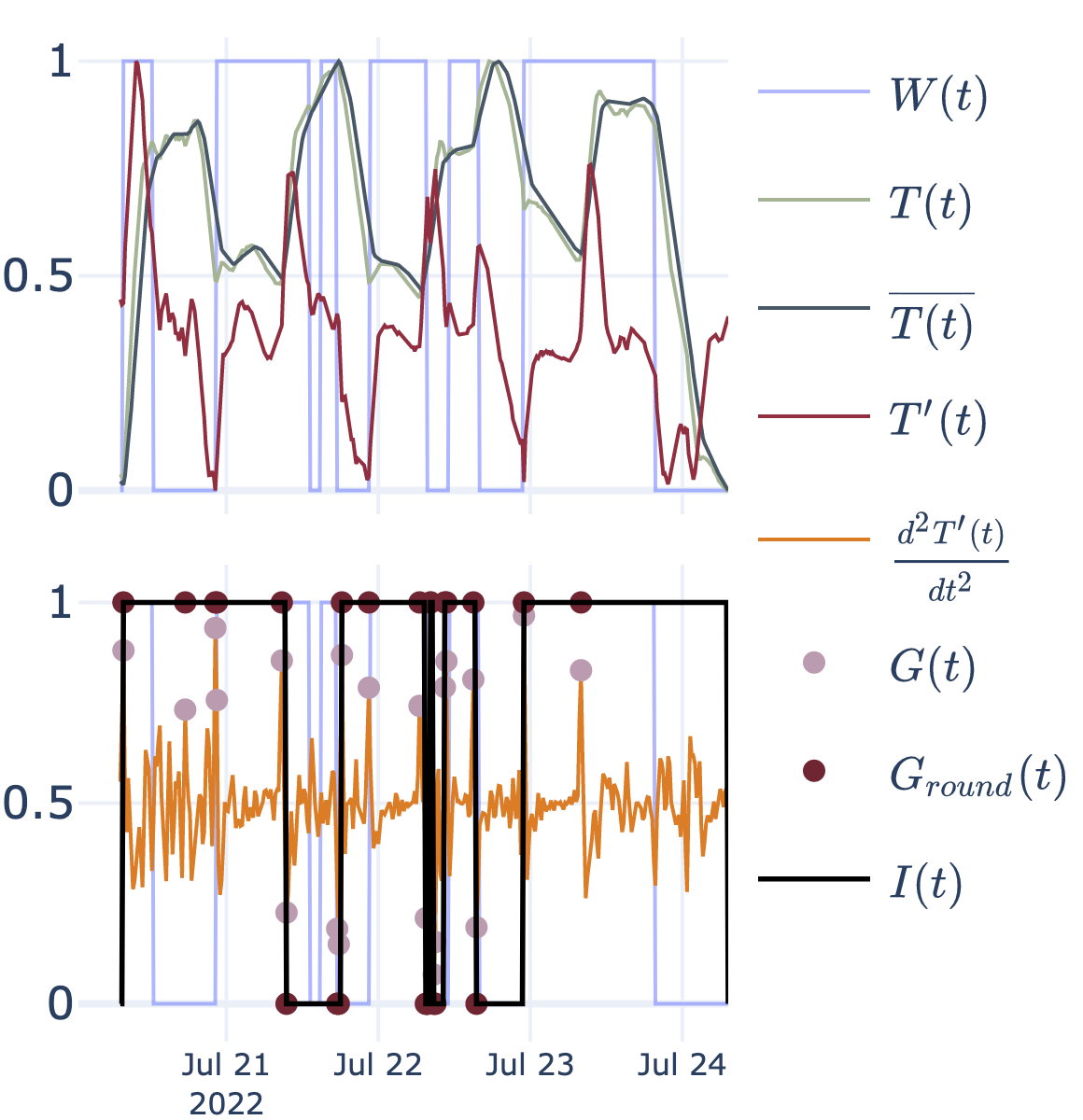}
\caption{Results of the Smoothing Technique}
\label{fig:st}
\end{figure}

The results of each step in ST are shown in Figure~\ref{fig:st}, where all time series have been normalized. The goal is to identify \(W(t)\), window state
as a function of time. This can take on two values: 0, representing
window closed, or 1, representing window open. We have an observed
variable \(T(t) = T_{meas}(t)\), which represents the measurement of the
indoor temperature. We apply an exponentially weighted mean (EWM)
function to \(T(t)\), creating a smoothed time series,
\(\overline{T(t)}\), which ideally removes strong peaks that would
reflect changes in window state, and isolates information concerning the
seasonal response and additional noise. This technique operates under
the assumption that the instantaneous change in indoor temperature due
to window opening is greater than any other potential source of
instantaneous temperature change. In reality, other unknown occurrences
within a room might cause large temperature spikes, which would
interfere with the efficacy of \(\overline{T(t)}\). Our choice of EWM came after preliminary analysis of alternative methods of smoothing, including seasonal trend decomposition  \cite{cleveland_cleveland_1990} and sinusoidal fit \cite{ollech_economic_2023}. Subtracting
\(\overline{T(t)}\) from \(T(t)\) yields \(T'(t)\), which is a time
series that reflects changes in the window state and some additional
noise.

In order to more confidently identify where the changes in window state
occur, we examine the first and second derivatives of \(T'(t)\),
\(\frac{dT'(t)}{dt}\) and \(\frac{d^2T'(t)}{dt^2}\). The second
derivative is particularly effective for identifying change points. In
order to predict where window changes occur, we apply a principle from
statistical hypothesis testing. We assume that the time series
\(\frac{d^2T'(t)}{dt^2}\) is normally distributed. Therefore, any value
in \(\frac{d^2T'(t)}{dt^2}\) that is more than 2 standard deviations
away from the mean of this time series is unlikely to occur, and could
possibly indicate an instance of a change in window state. We will use
these unlikely values as initial guesses \(G(t)\). They take on positive
or negative values depending on whether they are predicting a transition
from window open to close, or window close to open. Therefore, we round
the values of \(G(t)\) to 0 or 1 to reflect this. Finally, we
interpolate between the rounded values of \(G(t)\), so that we only
predict a change in window state when \(G(t)\) transitions between 0 and
1. This prediction of the window state is called \(I(t)\).

\hypertarget{machine-learning-method-support-vector-machine}{%
\subsubsection{Machine Learning (ML) Method: Support Vector
Machine}\label{machine-learning-method-support-vector-machine}}



Our focus was trying to get the best performance from the SVM using an
optimal set of features with optimal pre-processing functions applied.
We, therefore, developed a set of combinations of the various features we had
access to, as well as their derivatives and differences from one
another. The base features were: ambient temperature \(T_{amb}(t)\),
measured temperature \(T_{meas}(t)\), and their derivatives,
\(\frac{\mathrm{d}}{\mathrm{d}t}T_{meas}\) and
\(\frac{\mathrm{d}}{\mathrm{d}t}T_{amb}\). We also considered the
difference between ambient temperature and measured temperature
\(T_{amb} - T_{meas}\), the difference between measured temperature and
the derivative of measured temperature,
\(T_{meas} - \frac{\mathrm{d}}{\mathrm{d}t}T_{meas}\). We also included the
same features for relative humidity. After creating combinations of
these base features, we had a test set of 113 combinations.

For each data experiment, we created an SVM for each of
the combinations in the test set, which resulted in 3x113 different SVM
models.  As mentioned above, we are interested in
developing an unsupervised detection method. Therefore, we used the
One-Class SVM implementation from scikit-learn, which is ideal for anomaly detection.
Outliers are identified by clustering features to create the hyperplanes
for classification.


The SVM approach differs from ST that we have developed, and is similar
to other window techniques in that it is a generic ML method
that is not developed with window detection in mind. A wide array of
input features can be tested in order to get an acceptable prediction
accuracy. However, this might not be acceptable in practice, as
different scenarios or window operation behaviors, will demand different
sets of input features. In the event the true window detection pattern
is unknown, it will be difficult to know which input features are giving
a trustworthy and reliable result.

\hypertarget{evaluation-metrics}{%
\subsection{Evaluation Metrics}\label{evaluation-metrics}}

\hypertarget{de-rautlin-de-roy-metrics}{%
\subsubsection{De Rautlin de la Roy
Metrics}\label{de-rautlin-de-roy-metrics}} This set of metrics 
follow from a recent paper \cite{de_rautlin_de_la_roy_deep_2023} that compared the efficacy of different
machine learning algorithms for window detection, and reflect metrics that have been widely used in the literature.

\paragraph{Macro average F-1 score} The F-1 score is a classic metric for
evaluating the performance of classification algorithms. It condenses
information about the precision of a model in predicting a certain
class, as well as its ability to recall the available data. The
macro averaged F-1 score averages the F-1 scores for all individual
classes, but does not introduce weights in the averaging to reflect that
the classes may be unbalanced. As shown in
Table~\ref{tbl-data-collected}, the data sets we have vary from fairly
balanced in experiment A, to highly unbalanced in experiment B.
Therefore, using the macro average F-1 score provides a worst case
performance. Using this F1-score also provides a basis for comparison to
\cite{de_rautlin_de_la_roy_deep_2023}.

\paragraph{True or False Opening Time} This true opening time reflects the
total amount of time when \(I(t) = W(t) = 1\). False opening times occur
when \(I(t) = 1, W(t) = 0\). The metric is somewhat similar to the F-1
score in that rather than looking at specifically at change points, it
provides information the entire time period.

\hypertarget{custom-metrics}{%
\subsubsection{Custom Metrics}\label{custom-metrics}}

We developed this set of metrics based on the intuition that if a
model is perfectly able to capture the specific times when a window
state changes, then it is performing extremely well. We classify a
``guess'', as any value of \(I(t)\), and an ``action'' as any value of
\(W(t)\). A ``hit'' occurs when \(I(t) = W(t)\). Like \cite{de_rautlin_de_la_roy_deep_2023}, we consider
a prediction accurate if it is within at most 2 timesteps of the true
occurrence. Therefore, a ``near hit'' occurs when \(I(t) = W(t \pm 2)\).
The metrics we examine are given below.

\paragraph{Hits + Near Hits Over Guesses} This metric accounts for the
variability in guesses. A ratio of 1 indicates that all guesses taken by
the model were accurate within two timesteps. A ratio close to 0
indicates that a lot of guesses were taken, but relatively few were
close to where change occurred in the true window state.

\paragraph{Guesses Over Actions} This metric implicitly accounts for the
ability of the model to capture the pattern of the changes in window
state. If the true window state changed only 10 times, but the model
predicts changes on the order of 100, then the model is performing
poorly. A perfect score is 1.
\hypertarget{results}{%
\section{Results}\label{results}}

\begin{figure*}[t!] 
\begin{subfigure}{0.25\textwidth}
\includegraphics[width=\linewidth]{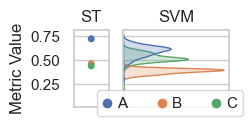}
\caption{Macro Average F1 Score} \label{fig:f1}
\end{subfigure}\hspace*{\fill}
\begin{subfigure}{0.22\textwidth}
\includegraphics[width=\linewidth]{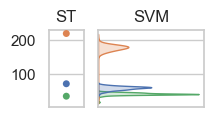}
\caption{True Opening Time [Hours]} \label{fig:tt}
\end{subfigure}\hspace*{\fill}
\begin{subfigure}{0.22\textwidth}
\includegraphics[width=\linewidth]{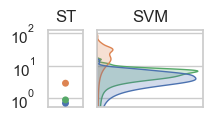}
\caption{Guesses/Actions} \label{fig:ga}
\end{subfigure}\hspace*{\fill}
\begin{subfigure}{0.22\textwidth}
\includegraphics[width=\linewidth]{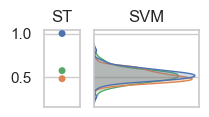}
\caption{Hits/Guesses} \label{fig:hg}
\end{subfigure}

\caption{Results across experiments and metrics (higher on y-axis is better performance, except (c) where score of 1 is optimal)} \label{fig:exp-results}
\end{figure*}

In Figure~\ref{fig:exp-results}, comparisons of ST to the performance of
the SVM across the metrics and experiments are displayed. While there was only one ST model created for each experiment, an SVM model was created for each 113 sets of features for each experiment. Therefore, the results for the SVM take on the form of a distribution. 

In Subfigure~\ref{fig:f1}, we see that the ST performs better than the
mean SVMs on experiments A and B, but lands closer to the bottom of the distribution for experiment C. The range of F1-scores displayed in
Subfigure~\ref{fig:f1} are comparable to those shown in \cite{de_rautlin_de_la_roy_deep_2023}. The
low values across all models for experiments B and C can be attributed
to the rather unbalanced nature of the datasets, which a macro average
F1-score penalizes.

The true time and false time metrics, shown in Subigure~\ref{fig:tt}
operates as an analog to a weighted average F1-score, in that the 
imbalance between window open and close is inherently taken into
account. The opening hours differ across experiments, with experiment B having windows open for about 330 hours, as shown in Table~\ref{tbl-data-collected}. Both the SVM and the ST underpredict this, but the SVM distribution has a mean of less than 200 as well as a long tail, indicating that the choice of features will significantly affect and SVM model's performance on these metrics. Windows are open for roughly 55 hours in experiments A and B. The ST and SVM both seem to overpredict opening times for experiment A, and underpredict closing times for experiment B.

For the more balanced experiment A, ST outperforms the SVM on the metric of hits/guesses Subfigure~\ref{fig:hg}. ST has similar accuracy for
experiments B and C compared to the mean of the SVM distributions.
 However, as Subfigure~\ref{fig:ga} highlights, the
SVMs take far more guesses per recorded action of window state change,
while ST takes on the order of 1 guess per recorded action. This
suggests that the SVMs' performance on the hits over guesses
metric is due to probability, rather than an embedding of the underlying
physical dynamics. This effect is most pronounced for experiment B, where the mean of the SVM distribution's guesses/actions is around 80. Although the ST value for the guesses/actions metric is high for experiment B as well, it is an order of magnitude lower than the SVM.

We see the most significant underperformance of the ST on experiment C, which is the most unfavorable dataset since it had unbalanced window states and narrow
distinction between indoor and outdoor temperature. Despite this, the ST shows
comparable or better performance compared the SVMs across all metrics for the other two datasets. This finding is significant, given that the SVMs shown in the
graphs have made unsupervised classification decisions on a wide range of
combinations of features, including ambient and indoor temperature and relative humidity. The ST method, which was developed using indoor temperature alone,
shows potential as a powerful technique with robust performance across
both favorable and unfavorable experimental datasets.
\hypertarget{conclusions-and-future-work}{%
\section{Conclusions and Future
Work}\label{conclusions-and-future-work}}

We presented preliminary results showcasing the efficacy of a new
method, ST, for unsupervised window detection. We compared these results
with SVM models trained with optimized features, and found comparable
results. Using datasets with differences in window state balance
and contrast between indoor and outdoor temperatures enabled us to begin
to test the limits of this new method. We also analyzed  the detection methods with a range of metrics that are better suited for unsupervised detection, as they reward models that are not only accurate on average, but also reflect the frequency and occurrence in time of window state change.



Although ST has shown promising results, we believe that its performance can be further improved. First, our method does not take into account the material properties of
the room where data is being collected, although this data would be
available from construction documentation. This information could help
to simulate response of indoor quantities of interest to daily seasonal
changes in ambient quantities using simple heat transfer equations. This
would enable us to more precisely identify the seasonal component in
indoor time series which would be an improvement from the simple
smoothing technique we use here. The addition of momentum equations to
our simulation would enable us examine the magnitude of temperature
change that convective heat transfer permits when a window is open vs
closed. Understanding expected indoor temperature changes under
different conditions would help to make more educated guesses in the
guessing step of the ST. In summary, augmenting the current data-driven
ST with a better understanding of the physical processes underlying the
data is a critical next step. The ST could also possibly be improved by carrying out a similar process
for relative humidity as temperature, as features involving relative
humidity featured heavily in the top-performing SVMs,


Spatial simulations and measurements could also be helpful in
determining where a sensor should be placed in a room to best capture
the fluctuations in quantities of interest. For this simulation, sensors
were intentionally placed away from from windows so as avoid drafts and
better capture the average temperature of a space. However, it is
conceivable that sensors placed closer to a window would be more
sensitive to changes in the window state, and thus record a stronger
window change detection signal.



\bibliographystyle{ACM-Reference-Format}
\bibliography{references}

\end{document}